\documentclass[aip, pof, reprint, twocolumn]{revtex4-1}
\usepackage{amsfonts,amsmath,bm,amssymb,color}
\usepackage{graphicx}
\usepackage[colorlinks=true, citecolor=blue, linkcolor=blue]{hyperref}

\begin{document}

\title{Inferring parameters and reconstruction of two-dimensional turbulent flows with physics-informed neural networks}

\author{Vladimir Parfenyev}\email{parfenius@gmail.com}
\affiliation{\mbox{Landau Institute for Theoretical Physics of the Russian Academy of Sciences, 142432 Chernogolovka, Russia}}
\affiliation{\mbox{HSE University, Faculty of Physics, 101000 Moscow, Russia}}

\author{Mark Blumenau}
\affiliation{\mbox{HSE University, Faculty of Physics, 101000 Moscow, Russia}}
\affiliation{\mbox{P.N. Lebedev Physical Institute of the Russian Academy of Sciences, 119991 Moscow, Russia}}

\author{Ilia Nikitin}
\affiliation{\mbox{Landau Institute for Theoretical Physics of the Russian Academy of Sciences, 142432 Chernogolovka, Russia}}
\affiliation{\mbox{HSE University, Faculty of Physics, 101000 Moscow, Russia}}

\date{\today}

\begin{abstract}
Obtaining system parameters and reconstructing the full flow state from limited velocity observations using conventional fluid dynamics solvers can be prohibitively expensive. Here we employ machine learning algorithms to overcome the challenge. As an example, we consider a moderately turbulent fluid flow, excited by a stationary force and described by a two-dimensional Navier-Stokes equation with linear bottom friction. Using dense in time, spatially sparse and probably noisy velocity data, we reconstruct the spatially dense velocity field, infer the pressure and driving force up to a harmonic function and its gradient, respectively, and determine the unknown fluid viscosity and friction coefficient. Both the root-mean-square errors of the reconstructions and their energy spectra are addressed. We study the dependence of these metrics on the degree of sparsity and noise in the velocity measurements. Our approach involves training a physics-informed neural network by minimizing the loss function, which penalizes deviations from the provided data and violations of the governing equations. The suggested technique extracts additional information from velocity measurements, potentially enhancing the capabilities of particle image/tracking velocimetry.
\end{abstract}

\maketitle

\section{Introduction} 

Experimental studies of fluid flows involve measuring their characteristics. The collected data may have low spatio-temporal resolution or be corrupted by measurement noise. Sometimes, the quantity of interest cannot be measured directly, and it has to be reconstructed from other measurements (e.g., pressure can be determined from a known velocity field~\cite{van2019pressure}). Various adjoint- and ensemble-variational data assimilation methods address these challenges by combining experimental data with the numerical simulations~\cite{zaki2021limited, asch2016data}. The basic idea is to find the flow state that satisfies the governing Navier-Stokes equation and minimizes the deviation from observations. The fundamental downside of such approaches is the computational complexity required to run multiple numerical simulations with various initial conditions and unknown system parameters that are adjusted during optimization. Once they are found, the related flow evolution gives access to the complete system state and at much higher resolution than the original data.

Recently, an alternative approach has been proposed in which the numerical solver is replaced by a physics-informed neural network (PINN)~\cite{raissi2019physics}. The flow estimation is also treated as a minimization problem. The neural network receives coordinates and a moment of time as input and returns flow variables as output. The loss function penalizes deviations from the available data and violations of the governing equations. Let us emphasize that the governing equations are no longer strictly obeyed; rather, their residuals are used as a penalty in the optimization problem. The main advantages of these methods are the relative simplicity of implementation and better suitability for ill-defined problems. In particular, PINNs may be used in domains with unknown boundary conditions~\cite{arzani2021uncovering}, whereas numerical simulations without that knowledge are impossible.

A side-by-side comparison of both methodologies was carried out in Ref.~[\onlinecite{du2023state}], where it was shown that PINNs are generally less accurate for sparse data. However, the situation may change in the future, since PINNs are a relatively new technology that is currently being actively improved. Examples of the application of this method to increasingly complex flow configurations can be found in recent reviews~\cite{cai2021physics,karniadakis2021physics, sharma2023review}.

In this paper, we apply PINNs to a specific example of two-dimensional turbulence, which is motivated by experimental studies~\cite{boffetta2005effects, xia2009spectrally, orlov2018large, fang2021spectral}. We assume that flow measurements reveal the velocity at some locations and time moments. These data can, for example, be obtained by particle image/tracking velocimetry (PIV/PTV) techniques, which track tracing particles over consecutive time frames to determine velocity. The collected data are assumed to be dense in time, but may be sparse in space and contain measurement errors. Density in time means that PIV/PTV methods are suitable for measuring instantaneous velocity with reasonable accuracy. Sparsity in space implies that we want to resolve scales that are small compared to the characteristic distance between measurement points. The data assimilation methods we have discussed connect successive moments in time by reproducing the dynamics determined by the Navier-Stokes equation. This allows information about the velocity field from different moments in time to be combined, increasing the accuracy and spatial resolution of possible reconstructions.

Our goal is to enhance the experimental data by increasing their spatial resolution and inferring flow variables and system parameters that were not measured directly. In particular, we aim to reconstruct the spatially dense velocity and pressure fields, establish the driving force, and unknown fluid viscosity and bottom friction coefficient, based only on velocity measurements. Knowledge of these additional quantities will allow us to better describe and characterize the properties of the considered system. The present investigation complements recent studies~\cite{eivazi2022aphysics, wang2022dense, clark2023reconstructing, cai2024physics} by exploring a different flow configuration.

The performed analysis shows that sparse ($150$ vectors per image) but accurate velocity measurements are sufficient to reconstruct a dense ($65536$ vectors per image) velocity field with a relative root-mean-square error of about $0.2 \%$ and similarly dense pressure and force fields with an accuracy several times worse for typical experimental conditions. The fluid viscosity and the bottom friction are restored with relative errors of several percent. The developed method is robust to small noise ($\leq 1\%$) in the initial velocity data and is even capable of correcting measurement errors based on the fact that the reconstructed quantities should satisfy the Navier-Stokes equation and incompressibility conditions. As the noise level in the initial data increases, the reconstruction accuracy gradually decreases. The changes in reconstruction accuracy with velocity data density are also addressed. Analysis of the energy spectra of velocity fields shows that small scales are the worst reconstructed, especially for noisy measurements.

\section{Governing equations}\label{sec:2}

We consider two-dimensional incompressible fluid flow described by the forced Navier-Stokes equation 
\begin{equation}\label{eq:1}
\partial_t \bm v + (\bm v \cdot \nabla) \bm v = - \nabla p -\alpha \bm v + \nu \nabla^2 \bm v + \bm f,
\end{equation}
where $\bm v$ is 2D velocity, $p$ is the pressure, $\alpha$ is the bottom friction coefficient, and $\nu$ is the kinematic viscosity. The external force $\bm f$ is assumed to be stationary in time. We are interested in a moderately turbulent regime, when the nonlinear term in the Navier-Stokes equation plays a substantial role. The flow is observed in some region with open boundaries, and to study the flow we measure the velocity field. We have no information about the pressure field and the external force, except that the latter is constant in time.

It is critical to recognize that in this formulation of the problem, pressure and external force cannot be clearly separated. Indeed, by performing the Helmholtz decomposition, the external force can be represented as the sum of the gradient and solenoidal fields. The gradient contribution can be included in the definition of pressure, and then, without loss of generality, the external force in equation (\ref{eq:1}) can be considered solenoidal, i.e. $\nabla \cdot \bm f = 0$. But even taking this condition into account, the division into pressure and external force is not unambiguous. These quantities are defined up to a transformation $p(\bm r,t) \to p (\bm r,t) + h(\bm r) + c(t)$, $f (\bm r) \to f (\bm r) + \nabla h(\bm r)$, where $h(\bm r)$ is a harmonic function $\nabla^2 h(\bm r) = 0$ in the observation zone and $c(t)$ is an arbitrary function that does not depend on spatial coordinates. An unambiguous definition of $h(\bm r)$ and $c(t)$ is impossible without knowledge of the boundary conditions for pressure and external force at the edges of the observation region. Note that $h(\bm r)$ does not depend on time, since the external force is assumed to be static.

The above should be kept in mind when determining the accuracy of field reconstructions. In the following, to characterize errors in the reconstructions of pressure and force fields, we will compare the values for their Laplacian $\nabla^2 p$ and curl $\phi = \partial_x f_y - \partial_y f_x$, respectively. Knowledge of velocity dynamics allows us to determine these quantities unambiguously.

\section{Data collection} 

\begin{figure}[t]
\centering{\includegraphics[width=\linewidth]{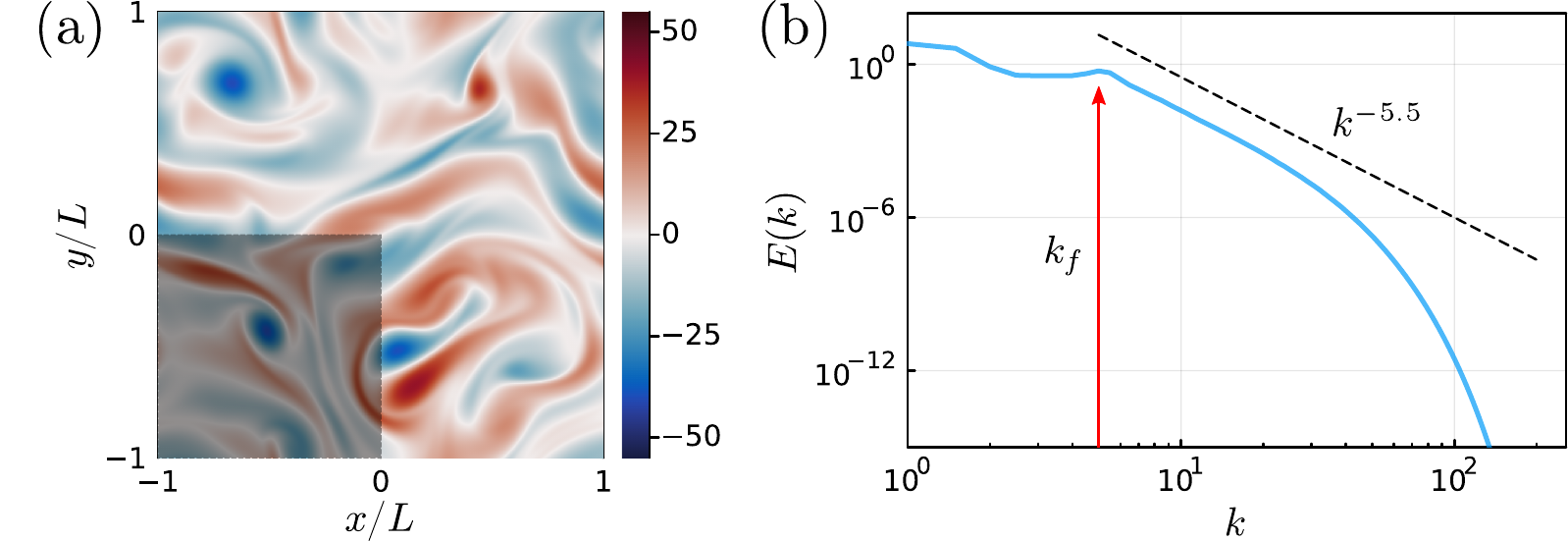}}
\caption{Vorticity field (a) and energy spectrum (b) in a statistical steady-state. The shaded area shows the observation zone.}
\label{fig:1}
\end{figure}

Data for this study are obtained numerically by integrating (\ref{eq:1}) using the GeophysicalFlows.jl pseudospectral code~\cite{constantinou2021geophysicalflows}, which has recently been successfully applied to model two-dimensional turbulence~\cite{parfenyev2022profile, kolokolov2024correlations, parfenyev2024statistical}. The code can be executed on the GPU, resulting in high computational performance. The aliasing errors are removed with the two-third rule. The simulation domain is a doubly periodic box of size $2 \pi \times 2 \pi$. The velocity measurements are performed in the observation area, which is smaller and has the size of $L \times L$ with $L=\pi$, see Fig.~\ref{fig:1}a. Since the boundary conditions at the edges of the observation zone are unknown, the application of adjoint- and ensemble-variational methods for data assimilation is problematic. 

The system parameters $\nu = 0.01$ and $\alpha=0.1$ are chosen to be consistent with the experimental studies~\cite{xia2009spectrally}. The external force has the Kolmogorov form
\begin{equation}\label{eq:2}
f_x = f_0 \sin (k_f y), \quad f_y = 0,
\end{equation}
with $f_0=10$ and $k_f=5$. We start from an arbitrary random initial condition, and after a transient process, the flow reaches a statistical steady-state. Once stationary, we saved data at $\Delta t = 0.02$ intervals for $T=4$ units, covering several turn-over times of flow fluctuations. This data recording rate corresponds to video recording in a laboratory experiment at $50$ frames per second. The spatial resolution of direct numerical simulation (DNS) is $512^2$ for the entire computational domain and $256^2$ for the observation zone.

Fig.~\ref{fig:1}a shows snapshot of the vorticity field $\omega = \partial_x v_y - \partial_y v_x$ in the statistical steady-state. The flow is chaotic, with no obvious imprint of an external forcing. Since there are two dissipation mechanisms -- viscosity and bottom friction -- the flow state can be characterized by two dimensionless parameters~\cite{mishra2015dynamics}: the Reynolds number $Re = UL/\nu \approx 1.3 \times 10^{3}$ and $Rh = U/(\alpha L) \approx 13$, which were computed using the root mean square velocity $U \approx 4.1$ and the size $L=\pi$ of the observation domain. The energy spectrum is presented in Fig.~\ref{fig:1}b. It indicates that fluid flow is mainly determined by modes with wave vectors $k \lesssim k_f$.

To imitate velocity measurements, we randomly scatter $N_{data}$ points $\{\bm r_n^d, t_n^d\}_{n=1}^{N_{data}}$ within the spatio-temporal ($256 \times 256 \times 200$) observation domain. We consider the case when the average distance between measurement points for a single moment in time is comparable to or larger than the forcing scale. It turns out that such sparse data are sufficient to perform reconstructions and infer system parameters with reasonable accuracy. In the experiments we rely on, the spatial resolution of the velocity measurements is usually much higher. In such a situation, it makes sense to thin out the measured velocity field by randomly selecting a relatively small number of data points, which can significantly speed up the reconstruction. The primary purpose in the above scenario is to infer the system parameters and flow variables that cannot be measured directly. If the experimental data are initially sparse in space, then the location of measurement points at close moments in time may have some spatial correlation that is not taken into account in our measurement model. The impact of such correlations on reconstruction accuracy requires further study and is beyond the scope of the present work.

The measured velocity $\bm v^d (\bm r_n^d, t_n^d)$ may differ from the DNS data $\bm v (\bm r_n^d, t_n^d)$ due to measurement noise. We model this distortion as
\begin{equation}
\bm v^d = 
\begin{pmatrix}
1+\eta_1 & 0\\
0 & 1+\eta_2
\end{pmatrix}
\bm v,
\end{equation}
where $\eta_i \sim \mathcal{N}(0, \varepsilon^2)$ are independent normally distributed random variables and the standard deviation $\varepsilon$ controls the noise level. Noise-free data correspond to $\varepsilon=0$. Note that this simple method to noise modeling accounts only for mistakes in estimating the velocity amplitude and ignores errors in wrongly referencing the velocity vector in space.

In most examples below, we set $N_{data} = 3 \times 10^4$ unless another value is explicitly specified. On average, this corresponds to $150$ measurement points per snapshot compared to $65536$ points at full resolution. The gathered data $\bm v^d (\bm r_n^d, t_n^d)$ are used for subsequent reconstruction of the flow state and system parameters with the PINN method.

\section{Physics-Informed Neural Network} 

\begin{figure}[t]
\centering{\includegraphics[width=\linewidth]{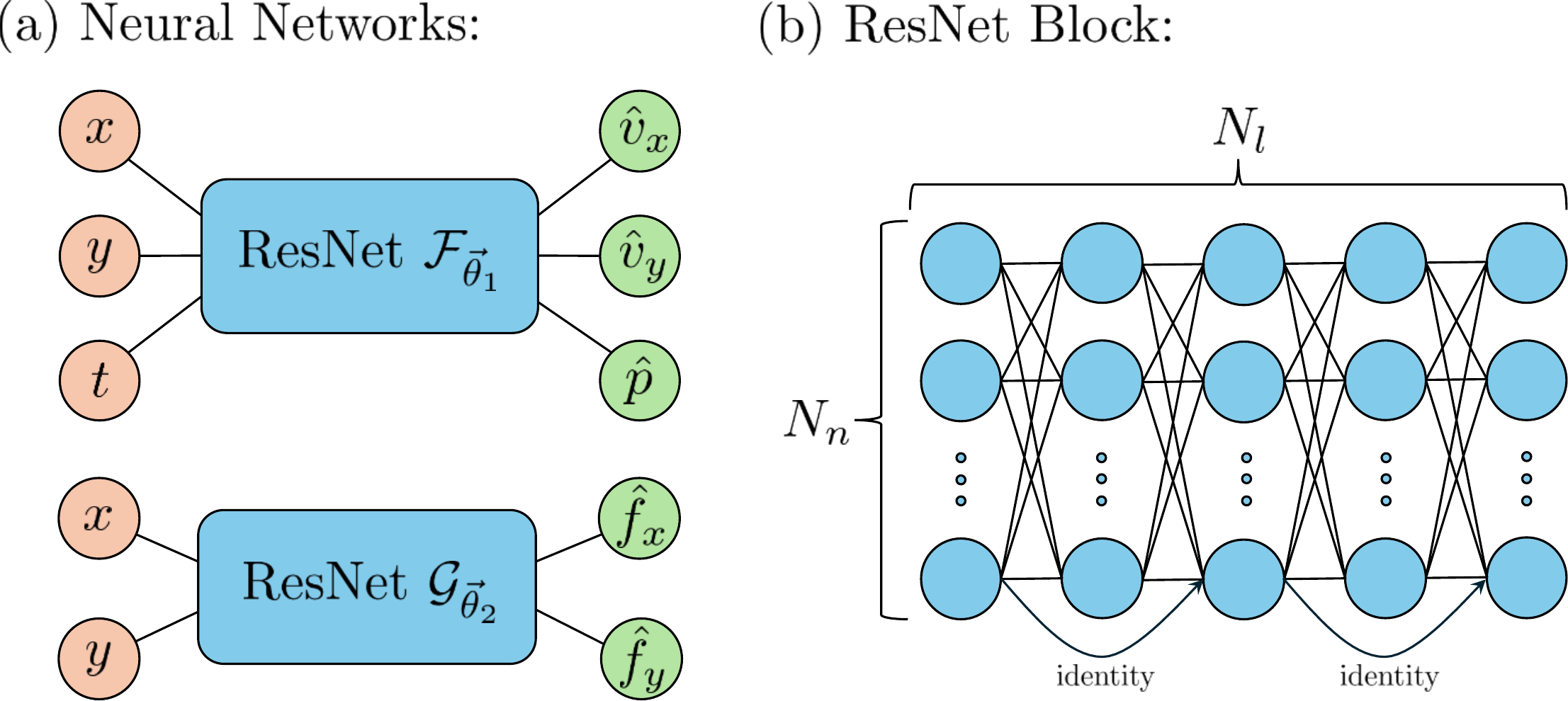}}
\caption{(a) Schematic of the neural networks and (b) adopted residual network architecture.}
\label{fig:2}
\end{figure}

In this section, we describe in detail the PINN method used in our study. We utilize two independent neural networks (NNs) that are trained simultaneously, see Fig.~\ref{fig:2}a. The first NN approximates the velocity and pressure at a particular position and time. It has three input neurons and three output neurons, $(\hat v_x, \hat v_y, \hat p) = \mathcal{F}_{\vec{\theta}_1}(x,y,t)$. The second NN is intended to predict stationary force. As a result, it contains only two input neurons and two output neurons, $(\hat f_x, \hat f_y) = \mathcal{G}_{\vec{\theta}_2}(x,y)$. The parameters $\vec{\theta}_1$ and $\vec \theta_2$ denote the trainable variables for NNs. In addition, we will assume that the bottom friction coefficient and the fluid kinematic viscosity are unknown to us. Since these parameters are positive, it is convenient to represent them in the form $\hat \alpha = e^{\lambda_1}$ and $\hat \nu = e^{\lambda_2}$. The scalar parameters $\lambda_1$ and $\lambda_2$ will be adjusted during PINN training along with $\vec{\theta}_1$ and $\vec \theta_2$.

\begin{figure*}[t]
\centering{\includegraphics[width=0.7\linewidth]{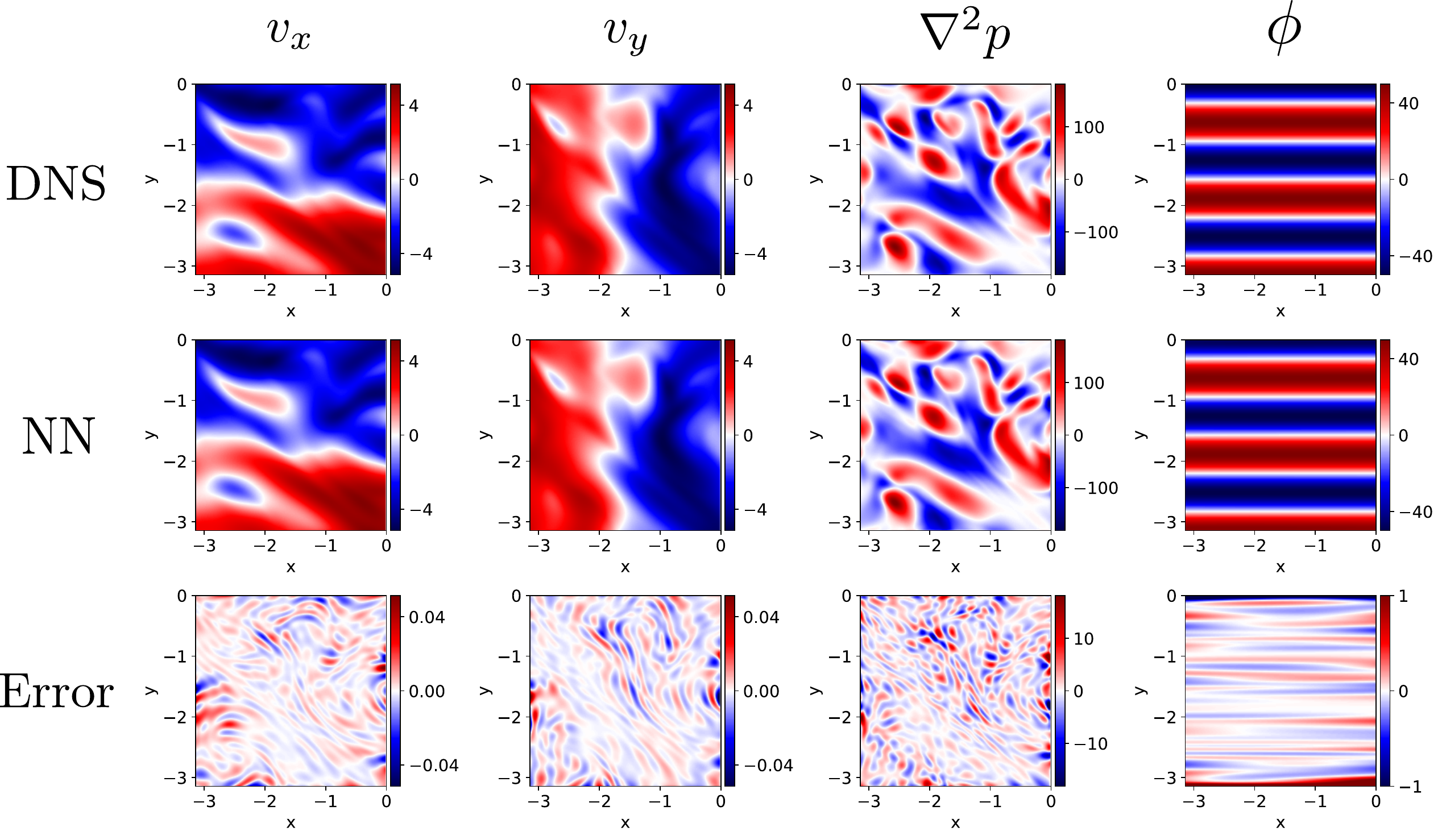}}
\caption{Comparison of DNS data (ground truth) and PINN predictions for a model trained on noise-free velocity measurements. The number of data points and collocation points are $N_{data} = 3 \times 10^4$ and $N_{eq}=9 \times 10^4$, respectively.}
\label{fig:3}
\end{figure*}

NNs must be sufficiently complicated to reproduce patterns that are essential for the considered turbulent flow. Universal approximation theorem ensures that this is possible~\cite{cybenko1989approximation, hornik1989multilayer}. In the original paper~\cite{raissi2019physics}, the authors used a fully connected NN architecture, but here we implement a residual design~\cite{he2016deep} for both NNs as we found that it performs better in agreement with the results reported earlier~\cite{cheng2021deep}. The adopted architecture is shown in Fig.~\ref{fig:2}b. The parameters $N_{l}$ and $N_{n}$ represent the number of hidden layers and neurons in each hidden layer, respectively. The output $\bm Y_k$ of $k$-layer is given by
\begin{eqnarray}
\nonumber
&\bm Y_{2j+1} = \sigma \left( \hat W_{2j+1} \bm Y_{2j} + \bm b_{2j+1} + \bm Y_{2j-1} \right), \quad j \geq 0,&\\
&\bm Y_{2j} = \sigma \left( \hat W_{2j} \bm Y_{2j-1} + \bm b_{2j} \right), \quad j \geq 1,&\\
\nonumber
&\bm Y_{N_l+1} = \hat W_{N_l+1} \bm Y_{N_l} + \bm b_{N_l+1},&
\end{eqnarray}
where $\bm Y_{-1} \equiv \bm 0$ for convenience of notation, $\bm Y_0$ corresponds to the input layer, $\bm Y_{N_l+1}$ denotes the output layer, $\hat W$ and $\bm b$ are weights and biases (trainable parameters that comprise $\vec \theta$), and $\sigma$ is the nonlinear activation function applied element-wise. We use a tanh activation function because it gives a reasonable trade-off between training time and resulting accuracy~\cite{wang2022dense}.

In general, increasing the size of NNs can reduce the prediction error, when the NNs are given enough data. Here we set $N_l=7$ and $N_n = 250$ for $\mathcal{F}_{\vec \theta_1}(\bm r,t)$, and $N_l=5$ and $N_n = 30$ for $\mathcal{G}_{\vec \theta_2} (\bm r)$. Note that $\mathcal{G}_{\vec \theta_2} (\bm r)$ is much simpler than $\mathcal{F}_{\vec \theta_1} (\bm r,t)$ because it is not time dependent and the external forcing usually does not have a complex spatial dependency.

\begin{figure*}[t]
\centering{\includegraphics[width=0.9\linewidth]{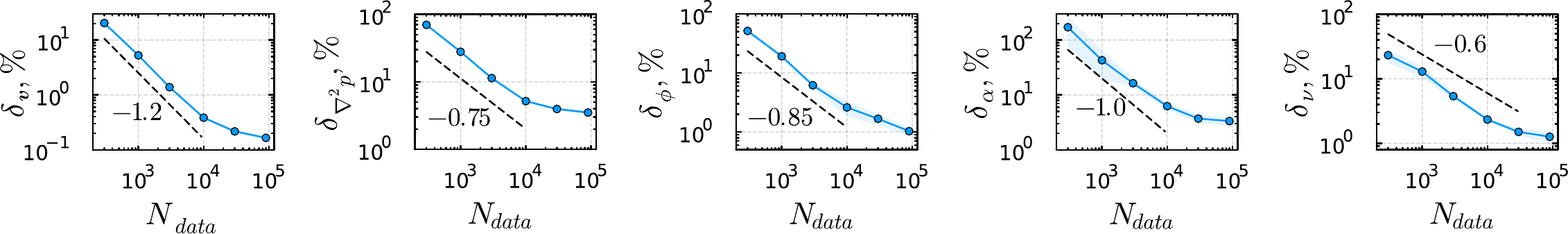}}
\caption{Dependence of relative root-mean-square errors of PINN reconstructions on the number $N_{data}$ of data points used for training. Dashed lines are added for reference and their slopes are indicated. The shaded areas show the standard deviations that were estimated by training five independent PINN models for each value of $N_{data}$.}
\label{fig:4}
\end{figure*}

Training NNs involves minimizing a loss function consisting of two terms. The first term represents the deviation between predicted and measured velocity
\begin{equation}
\mathcal{L}_{data} = \dfrac{1}{N_{data} U^2} \sum_{n=1}^{N_{data}} |\hat{\bm v}(\bm r_n^d, t_n^d) - \bm v^{d} (\bm r_n^d, t_n^d)|^2,
\end{equation}
where $U^2 = \sum_n \bm v_d^2 (\bm r_n^d, t_n^d)/N_{data}$ is measurement-based estimate of mean squared velocity. The second term penalizes deviations of predictions from the Navier-Stokes equation and incompressibility conditions
\begin{equation}
\mathcal{L}_{eq} = \dfrac{1}{N_{eq}} \sum_{n=1}^{N_{eq}} \left[ \bm e_1^2 (\bm r_n^e, t_n^e) + e_2^2 (\bm r_n^e, t_n^e) + e_3^2 (\bm r_n^e) \right],
\end{equation}
where $\{\bm r_n^e, t_n^e\}_{n=1}^{N_{eq}}$ are the collocation points at which equation loss is calculated and
\begin{eqnarray*}
&\bm e_1 = \left(\partial_t \hat{\bm v} + (\hat{\bm v} \cdot \nabla) \hat{\bm v} + \nabla \hat{p} + e^{\lambda_1} \hat{\bm v} - e^{\lambda_2} \nabla^2 \hat{\bm v} - \hat{\bm f} \right)L/U^2, \;&\\
&e_2 = \left(\partial_x \hat v_x + \partial_y \hat v_y \right)L/U,&\\
&e_3 = \left(\partial_x \hat f_x + \partial_y \hat f_y \right) L^2/U^2.&
\end{eqnarray*}
Computing the right-hand sides involves differentiating the outputs of NNs with respect to input neurons using automatic differentiation~\cite{baydin2018automatic}. The number and location of collocation points are not limited by measurements and can be chosen arbitrarily. We set $N_{eq}=9 \times 10^4$, with part of these points corresponding to data points $\{\bm r_n^d, t_n^d\}_{n=1}^{N_{data}}$, and the rest are generated randomly. Note that instead of requiring the incompressibility conditions to be satisfied, the NNs can be modified to predict the relevant stream functions~\cite{raissi2019physics, kag2022physics}. However, in this case, calculating the terms in the Navier-Stokes equation involves computation of higher-order derivatives, which significantly slows down the training process.

Finally, the total loss function is given by
\begin{equation}\label{eq:loss}
  \mathcal{L} = \mathcal{L}_{data} + \beta \mathcal{L}_{eq},
\end{equation}
where $\beta$ is a weighting coefficient used to balance two terms. Its value can be adaptively changed during training~\cite{wang2021understanding, jin2021nsfnets}, but here we use a simpler strategy in which the weight is fixed. Empirically, we have found that $\beta=0.02$ is well suited for this type of problem.

Minimization of the loss function (\ref{eq:loss}) is performed by a gradient-based optimizer that adjusts network $\{\vec \theta_1, \vec \theta_2 \}$ and system $\{ \lambda_1, \lambda_2 \}$ parameters. First, we train the model for $10^5$ epochs using Adam optimizer~\cite{kingma2014adam} with learning rate of $\mu = 10^{-3}$. An epoch refers to one complete pass of the training dataset. To reduce computation time, we have implemented a mixed precision technique~\cite{micikevicius2017mixed}. Then we use L-BFGS-B optimizer~\cite{byrd1995limited} until the tolerance reaches machine precision. The training is carried out on a single NVIDIA A100 GPU. The code is written using the PyTorch library~\cite{paszke2019pytorch} and is openly available at~[\onlinecite{repository}].

\section{Results}

To quantify the performance of trained models, we introduce relative root-mean-square errors (RRMSEs), which involve averaging both over observation zone $||\dots||_2 = \sqrt{\frac{1}{L^2} \int d^2 r |\dots|^2}$ and across all time points $\langle \dots \rangle_T = \frac{1}{T} \int_0^T dt \dots$
\begin{equation}
\delta_Q = \left\langle \dfrac{||\hat Q - Q||_2}{||Q||_2} \right\rangle_T,
\end{equation}
where $\hat Q$ denotes the physical quantity of interest predicted by PINN, and $Q$ is its exact value from DNS. We will look at how well velocity, pressure, external force, bottom friction, and fluid viscosity are reconstructed. However, as discussed in Section~\ref{sec:2}, the pressure and external force cannot be determined unambiguously from velocity field measurements. Therefore, to determine the accuracy of reconstructions of these fields, we will analyze errors in determining the pressure Laplacian $\nabla^2 p$ and the external force curl $\phi = \partial_x f_y - \partial_y f_x$. Furthermore, to test the statistical properties of the predicted velocity, we will compare the energy spectra of the reference and reconstructed flows.

\begin{table*}[t]
\caption{Dependence of mean values and standard deviations for prediction errors on the noise level $\varepsilon$.}
\label{tab:1}
\begin{ruledtabular}
\begin{tabular}{c|c|c|c|c|c|c}
         $\varepsilon$   & $\delta_{\bm v}, \%$ & $\delta_{\nabla^2 p}, \%$ &  $\delta_{\phi}, \%$  &  $\delta_{\alpha}, \%$ & $\delta_{\nu}, \%$ & $\delta_{p-\langle p \rangle_T}, \%$ \\ \hline
         $0$ & $0.22 \pm 0.01$ & $4.0 \pm 0.2$ &  $1.6 \pm 0.3$  &  $3.9 \pm 0.5$ & $1.5 \pm 0.1$ & $0.58 \pm 0.03$ \\
         $0.005$ & $0.28 \pm 0.01$ & $4.7 \pm 0.1$ &  $1.7 \pm 0.4$  &  $3.1 \pm 0.9$ & $1.2 \pm 0.2$ & $0.86 \pm 0.03$ \\
         $0.01$ & $0.50 \pm 0.03$ & $10.1 \pm 1.8$ &  $2.0 \pm 0.4$  &  $2.2 \pm 1.8$ & $0.7 \pm 0.5$ & $1.7 \pm 0.1$ \\
         $0.02$ & $2.39 \pm 0.06$ & $157 \pm 9$ &  $5.4 \pm 0.8$  &  $41.4 \pm 4.4$ & $12.3 \pm 0.9$ & $7.7 \pm 0.3$ \\
         $0.05$ & $5.84 \pm 0.28$ & $478 \pm 43$ &  $13.0 \pm 2.7$  &  $188 \pm 26$ & $56.1 \pm 2.0$ & $17.3 \pm 1.0$
\end{tabular}
\end{ruledtabular}
\end{table*}


\subsection{Impact of input data density}

First, we trained PINN models based on noise-free input data. We set a fixed number $N_{eq}=9 \times 10^4$ of collocation points and vary the number $N_{data}$ of data points in the range from $3 \times 10^2$ to $9 \times 10^4$ to study the dependence of prediction accuracy on input data density. Reconstructions employing input data with $N_{data} \geq 10^4$ are highly accurate. Fig.~\ref{fig:3} compares DNS data with NN predictions in the observation zone at some instant of time for a PINN model trained with $N_{data} = 3 \times 10^4$ measurement points. The outcomes appear to be very close, with differences mostly localized on small scales.

Fig.~\ref{fig:4} reports RRMSEs for determining the velocity, pressure Laplacian, external force curl, bottom friction, and fluid viscosity. As $N_{data}$ increases, RRMSEs first decrease rapidly in a power-law manner ($N_{data} \leq 10^4$), and then the dependence gradually begins to saturate ($N_{data} \geq 10^4$). Similar behavior was observed earlier for the simpler problem of reconstructing 2D Taylor's decaying vortices~\cite{wang2022dense}. The velocity reconstruction error is less than $0.5\%$ for $N_{data} \geq 10^4$, while for other quantities the errors are approximately an order of magnitude larger. This behavior is due to the lack of measurements for these quantities, so the model extracts them solely from the analysis of governing equations.

\begin{figure}[b]
\centering{\includegraphics[width=0.7\linewidth]{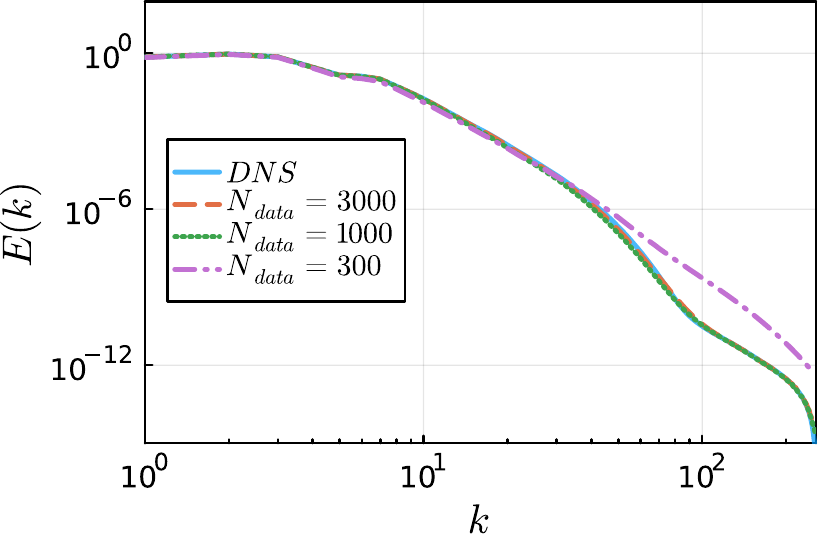}}
\caption{Energy spectra for DNS data and PINN models trained with different numbers $N_{data}$ of data points.}
\label{fig:5}
\end{figure}

In Fig.~\ref{fig:5}, we compare the energy spectra of the reference flow and PINN reconstructions, which are commonly used to characterize the statistical properties of a velocity field. Since the observation region is not periodic, when constructing the spectra we applied a Hann window function. For this reason, the DNS spectrum is slightly different from that presented earlier in Fig.~\ref{fig:2}b. The statistical properties of the velocity field are well reproduced for $N_{data} \geq 10^3$. Any noticeable differences are observed only for more sparse data, and they are mainly localized at small scales with $k \gg k_f$.

\subsection{Noisy velocity measurements}

Next, we fix $N_{data} = 3 \times 10^4$ and repeat the training, but now the models receive noisy measurement data. Table~\ref{tab:1} reports the dependence of the average values and standard deviations for prediction errors on the noise level, obtained by averaging over ten models in each case. The developed algorithm is robust to small noise ($\varepsilon \leq 0.01$) in the input data. Moreover, the error for the reconstructed velocity in this case is less than the noise level. Thus, PINN is able to correct for small noise based on the knowledge that the velocity field must satisfy the Navier-Stokes equation and incompressibility condition. 

As the noise level increases ($\varepsilon \geq 0.02$), the performance of the trained models gradually decreases. Since for a turbulent flow the dissipative terms make a relatively small contribution to the Navier-Stokes equation, the model quickly loses the ability to predict the bottom friction $\alpha$ and fluid viscosity $\nu$. At the same time, the accuracy of the velocity field and external force determination remains comparable to the noise level. 

\begin{figure}[b]
\centering{\includegraphics[width=0.7\linewidth]{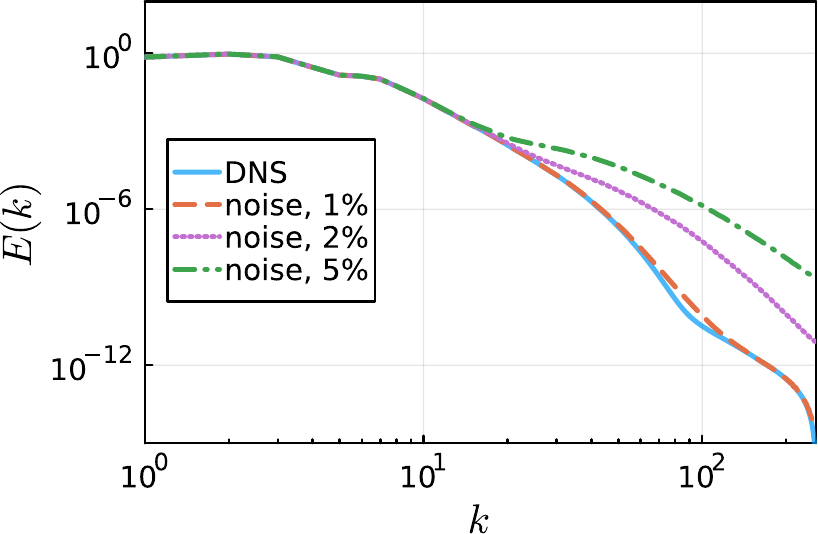}}
\caption{Energy spectra for DNS data and PINN models trained with various noise levels in the input data.}
\label{fig:6}
\end{figure}

Errors in predicting the pressure Laplacian take on gigantic values, but in fact this is due to the peculiarities of the introduced metric, and the pressure field itself is reconstructed quite well. The presence of noise in the training data leads to small-scale errors, which are amplified when calculating the second-order spatial derivatives. A similar statement applies to the interpretation of errors in determining the external force curl, but to a lesser extent, since in this case only first-order derivatives are calculated. Fig.~\ref{fig:6} compares the energy spectra for the reference and reconstructed flows, and supports the statement that there are small-scale errors in the predicted velocity. 

\begin{figure}[b]
\centering{\includegraphics[width=0.9\linewidth]{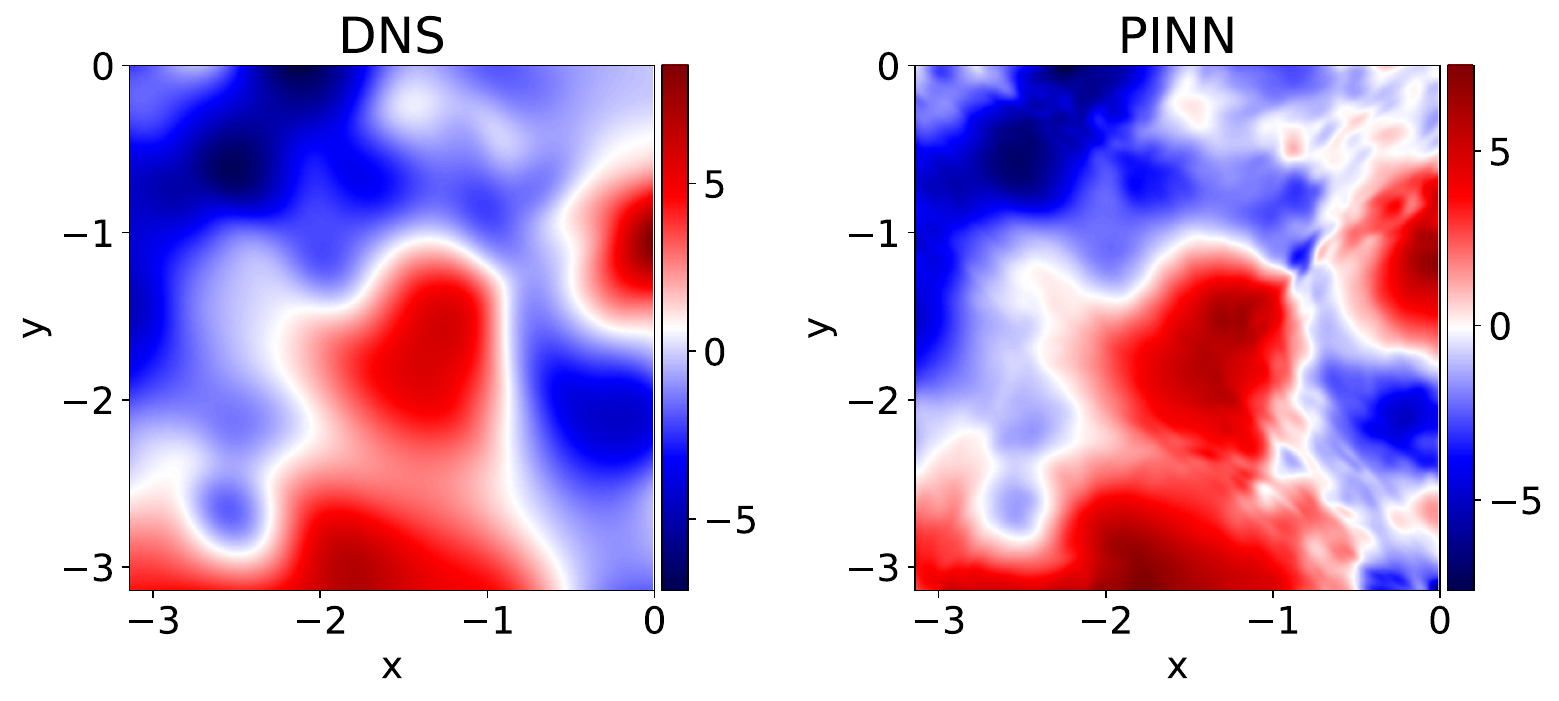}}
\caption{Comparison of modified pressure $p - \langle p \rangle_T$ for DNS data and a PINN model trained on noisy input data with $\varepsilon=0.05$.}
\label{fig:7}
\end{figure}

To evaluate the accuracy of pressure reconstruction in a more direct way, we consider the modified pressure $p-\langle p \rangle_T$. Subtracting the mean value over time allows one to eliminate the ambiguity in determining pressure associated with an arbitrary harmonic function $h(\bm r)$, see details in Section~\ref{sec:2}. The modified pressure is determined up to an arbitrary constant, so when comparing these fields we also subtract the spatial average values from them. The results are illustrated in Fig.~\ref{fig:7}, where we compare the modified pressures at some moment of time for DNS data and a PINN model trained on velocity measurements with a noise level of $\varepsilon = 0.05$. Despite the noticeable differences, the modified pressures are generally similar to each other. The corresponding RRMSEs for all noise levels are shown in the last column of Table~\ref{tab:1} and they are significantly smaller than the errors in the pressure Laplacian reconstruction.

\section{Conclusion}

We showed that PINN can combine dense in time, spatially sparse and probably noisy velocity measurements with an accurate physical model to extract additional information from ordinary flow observations. The developed technique can be built on top of current PIV/PTV workflows, potentially improving their functionality. In particular, we investigated a forced two-dimensional turbulent fluid flow and demonstrated that the proposed method can reconstruct the spatially dense velocity and pressure fields, infer the driving force, and determine the unknown fluid viscosity and bottom friction coefficient. The ability to reconstruct the full flow state from limited observations can also be used to compress experimental data.

The suggested technique is based on training NNs by minimizing a loss function that penalizes deviations from measured data and violations of the governing equations. While our intention was to select the optimal NN design and hyperparameter configuration, a thorough examination of this issue was outside the scope of this work. We anticipate that the introduction of more advanced NN architectures~\cite{zhao2023pinnsformer, wang2024piratenets, liu2024kan, Buzaev2023fourrieretc} can further improve PINN method. 

The reconstructed flows were compared to the reference flow in terms of relative root-mean-square errors and energy spectra. Overall, the results show good accuracy for super-resolution and inference applications and demonstrate moderate robustness to noise in the input data. At low noise levels, the algorithm can correct errors in the velocity measurements based on governing equations. The developed technique can be useful for analyzing flow observations when the physical model of the phenomenon is well constructed.

Although our study used time-dense velocity measurements, we would like to point out that the PINN method can be potentially extended and applied to time-sparse measurements as well. In this case, the interval between successive images of tracer particles is long enough and PIV/PTV methods are ineffective for determining instantaneous velocities. However, it is possible to complement the physical model with an equation describing the advection of tracer particles, and then train PINN on their experimental images. Examples of this strategy can be found in Ref.~[\onlinecite{raissi2020hidden}], and instead of tracer particles, it is also possible to monitor the temperature field~\cite{cai2021flow, biferale2023reconstructing}. For turbulent systems, the distance between successive data frames cannot exceed the Lyapunov time, but how close one can get to this limit is an open question.

Finally, while the presented results look promising, the study did not address a number of issues that may be relevant for practical applications:
\begin{itemize}
    \item In many cases, the adopted physical model only approximately describes the system. For flows in thin fluid layers, equation (\ref{eq:1}) is only an approximation, and it breaks down for small-scale modes whose size is comparable to the fluid layer thickness. How sensitive is the PINN algorithm to modeling flaws?
    
    \item The flow under consideration was spatially uniform. Will the algorithm work equally well for substantially non-uniform systems? What modifications would it require?

    \item The procedure used to model the velocity measurements did not take into account all the features of experimental PIV/PTV measurements. It would be helpful to investigate the algorithm's capabilities using experimental data collected under various settings. The first steps in this direction were done in Refs.~[\onlinecite{wang2022dense, clark2023reconstructing, cai2024physics}].

    \item How increasing the degree of flow non-linearity affect the performance of the algorithm? And how much data are required to achieve the desired reconstruction accuracy? Is it possible to approach the problem theoretically and find fundamental limitations on the required spatio-temporal data density?
\end{itemize}
We hope that future research will address these challenges and that the solutions will contribute to the further development of PINN.

\acknowledgments

The work of VP and IN was supported by the Ministry of Science and Higher Education of the Russian Federation within the State Assignment No. FFWR-2024-0017 of Landau Institute for Theoretical Physics. VP also acknowledges support from the Foundation for the Advancement of Theoretical Physics and Mathematics ''BASIS'', Project No. 22-1-3-24-1. The authors are grateful to the Landau Institute for Theoretical Physics for providing computing resources.

\section*{Data AVAILABILITY}
The data that support the findings of this study are openly available at \url{https://github.com/parfenyev/2d-turb-PINN/}.

\bibliography{biblio}

\end{document}